\documentclass[a4paper,12pt]{article}
\usepackage[latin1]{inputenc}
\usepackage{amsmath}
\usepackage{amssymb}
\usepackage{graphicx}
\begin{document}

\title{\textbf{``Luttinger'' and insulating spin liquids in two dimensions}}
\author{H. Freire, E. Corrêa and A. Ferraz \\ \\
\small\emph{Laboratório de Supercondutividade,}\\
\small\emph{Centro Internacional de Física da Matéria Condensada,}\\
\small\emph{Universidade de Brasília - Brasília, Brazil}}
\date{}
\maketitle

\begin{abstract}
In the present work, we implement an explicit two-loop renormalization
of a two-dimensional flat Fermi surface (FS) in the framework of a
field theoretical renormalization group (RG) approach. In our scheme,
we derive the RG equations for both coupling functions and Fermi energy.
In this way, we are able to probe the existence of spin-charge separation
by showing that the low-energy sector of the system is in fact a non-Fermi
liquid. In addition, associating the true interacting FS to the infrared
stable (IR) fixed point of the Fermi energy, we demonstrate here that
it either acquires a small curvature and behaves as a {}``Luttinger
liquid'' or it suffers a truncation in k-space depicting an insulating
spin liquid.
\end{abstract}

\section*{1 - Introduction}

A better understanding of the physical properties of highly interacting
electrons in two spatial dimensions (2d) is central for high-Tc superconductivity.
Soon after the discovery of the high-Tc superconductors, Anderson~\cite{Anderson}
suggested that a strongly interacting 2d electron gas should resemble
a 1d Luttinger liquid state. This question remains unresolved to this
date. Thanks to the high precision of the angular resolved photoemission
experiments performed in a variety of materials~\cite{Campuzano},
we know, at present, important facts concerning the Fermi surface
(FS) of the cuprates. The FS's for underdoped and optimally doped
Bi2212 and YBaCuO compounds contain both flat and curved sectors~\cite{Dessau}.
As a result, they are nearly perfectly nested along certain k-directions.
As is well-known, whenever there is a flat FS, the corresponding one-electron
dispersion is 1d-like in momentum space.

Originally, the cuprates are Mott insulators which become metallic
at very low doping~\cite{Yoshida}. At half-filling, Hubbard-like
models have a square shape FS imposed by electron-hole symmetry.
The FS changes as we vary the filling factor and, as soon as it is
lightly doped, it acquires nonzero curvature sectors in k-space.
In the immediate vicinity of half-filling, there are, at most,
isolated curved spots in momentum space. Consequently, in a zeroth
order approximation, one may neglect their presence altogether.
Following this scheme, several workers investigated the properties
of a 2d electron gas in the presence of a totally flat
FS~\cite{Ruvalds,Dzyaloshinskii,Luther,Sudbo}. In their
approaches, the FS is always kept fixed and never deviates from
its original flat form. Besides that, their results conflict with
each other. Conventional perturbation theory
calculations~\cite{Ruvalds}, parquet method
results~\cite{Dzyaloshinskii}, as well as one-loop perturbative RG
calculations~\cite{Doucot} indicate that, for repulsive
interactions, there is never a Luttinger liquid state in 2d. In
contrast, applying bosonization methods, Luther was able to map
the square FS onto two sets of perpendicular chains~\cite{Luther}.
As a result of that, the corresponding electron correlation
functions become sums of power law terms with exponents only
differing in form from those of a Luttinger liquid~\cite{Sudbo}.
We revisit this problem in this letter.

\begin{figure}
\centering
\includegraphics[height=2.2in]{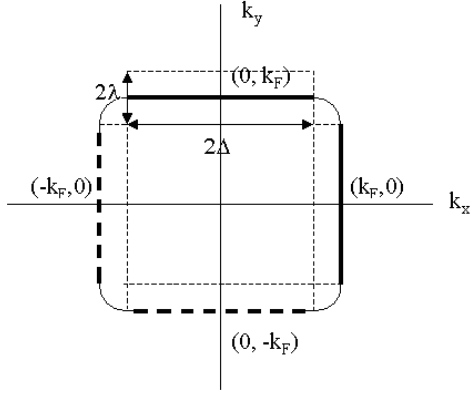}
\caption{\small The initial Fermi surface. The corners are rounded
to avoid van Hove singularities.} \label{f.1}
\end{figure}

We report a two-loop field theoretical RG calculation for the electron
gas in the presence of the same FS model as used by Dzyaloshinskii
and co-workers~\cite{Dzyaloshinskii}. The novel aspect of our work,
apart from taking into account important higher order corrections,
is the fact that we show explicitly how the FS changes its shape due
to interactions. As a result, we are able to determine when the FS
may suffer a truncation in k-space. To our knowledge, this was never
done before in such a systematic and detailed form. Needless to say,
to show how the FS is renormalized by interaction is a very intricate
many-body problem~\cite{Nozieres}.

To deal with this here, we calculate how the charge
renormalization functions Z and all physical parameter vary along
the renormalized FS itself, by means of appropriate RG flow
equations for the Fermi energy and coupling functions. We observe
that Z is nullified at FS and, as a result, there exists
spin-charge separation in 2d. In addition, we explore the
existence of nontrivial fixed points which vary continuously along
the Fermi surface. Accordingly, we show that the renormalized FS
naturally develops a nonzero curvature, if one associates its
physical nature to the infrared (IR) stable fixed points. It
follows from this that its behavior is regulated by the variation
of the anomalous dimension exponent $\gamma ^{*}$ with respect to
the momentum $p_{||}$ along FS. When $1/2\leq \gamma ^{*}\leq 1$
the renormalized FS is truncated and develops a charge pseudogap
in this region of k-space. The physical system behaves as an
insulating spin liquid in this case. In contrast, when $0<\gamma
^{*}<1/2$ the non-Fermi liquid is metallic and resembles a
Luttinger liquid state.

\section*{2 - Renormalized hamiltonian and electron self-energy}

Our starting point is a strongly interacting 2d electron gas in
the presence of the flat FS shown schematically in fig.~\ref{f.1}.
In order to keep a closer contact with well-known works in
one-dimensional physics~\cite{Solyom}, we split the FS in four
patches. However, we expand the bare single-particle energy
dispersion in the vicinity of the renormalized (i.e physical) FS.
The parametrization of the corresponding interactions is shown
schematically in fig.2. If we use this model to calculate physical
quantities using a naive perturbation theory, we find divergent
results for particular values of the external
momenta~\cite{Ferraz1,Ferraz2}. We circumvent this problem
following the standard field theory procedure of introducing
appropriate counterterms in the hamiltonian to render the physical
parameters finite in all scattering channels~\cite{Peskin}. In
this way, the original hamiltonian is rewritten in a more
convenient form in terms of the low-energy parameters, which are
in turn physically measurable. Thus, we have $H=H_{R}+H_{C}$,
where

\begin{eqnarray*}
H_{R}=\sum _{\mathbf{p}\sigma }v_{FR}(|p_{\perp }|-k_{FR})\psi _{R\sigma }^{\dagger }(\mathbf{p})\psi _{R\sigma }(\mathbf{p}) &  &
\end{eqnarray*}

\begin{equation}
+\sum _{\mathbf{pqk}\sigma }(U_{1R}+U_{2R}+U_{3R}+U_{4R})\psi _{R\sigma }^{\dagger }(\mathbf{p+q-k})\psi _{R,-\sigma }^{\dagger }(\mathbf{k})\psi _{R,-\sigma }(\mathbf{q})\psi _{R,\sigma }(\mathbf{p})\label{e.1}\end{equation}
\\
 and \begin{eqnarray*}
H_{C}=\sum _{\mathbf{p}\sigma }[\frac{Z(\mathbf{p})}{2m_{B}}(k_{FR}^{2}-k_{FB}^{2})+(Z(\mathbf{p})\frac{m_{R}}{m_{B}}-1)v_{FR}(|p_{\perp }|-k_{FR})]\psi _{R\sigma }^{\dagger }(\mathbf{p})\psi _{R\sigma }(\mathbf{p}) &  &
\end{eqnarray*}

\begin{equation}
+\sum _{\mathbf{pqk}\sigma \sigma ^{'}}(\Delta U_{1R}^{\sigma
\sigma ^{'}}+\Delta U_{2R}^{\sigma \sigma ^{'}}+\Delta
U_{3R}^{\sigma \sigma ^{'}}+\Delta U_{4R}^{\sigma \sigma
^{'}})\psi _{R\sigma }^{\dagger }(\mathbf{p+q-k})\psi _{R,\sigma
^{'}}^{\dagger }(\mathbf{k})\psi _{R,\sigma ^{'}}(\mathbf{q})\psi
_{R,\sigma }(\mathbf{p})\label{e.1}
\end{equation}
Here the subscripts ``R'' and ``B'' stand for renormalized and
bare respectively. Besides, $Z(\mathbf{p})$ is the charge
renormalization function, which is well-defined only in the
following regions of k-space depicted in fig.1: $-\lambda \leq
p_{\perp }\mp k_{FR}\leq \lambda $ and $-\Delta \leq p_{\Vert}\leq
\Delta $. Finally, the renormalized and bare fields are related to
each other by $\psi _{B}(\mathbf{p})=Z^{1/2}(\mathbf{p})\psi
_{R}(\mathbf{p})$, whereas the renormalized and bare couplings are
connected by
$\prod_{i=1}^{4}Z^{1/2}(\mathbf{p}_{i})U_{iB}=U_{iR}+\Delta
U_{iR}^{\sigma \sigma ^{'}}$. Even if, initially, the couplings
are taken to be constants, the renormalization process necessarily
forces $U_{iR}$ to be momenta dependent all along $FS$. The
counterterms, for this reason, form a continuum set in momenta
space.

\begin{figure}
\centering \includegraphics[height=1.2in]{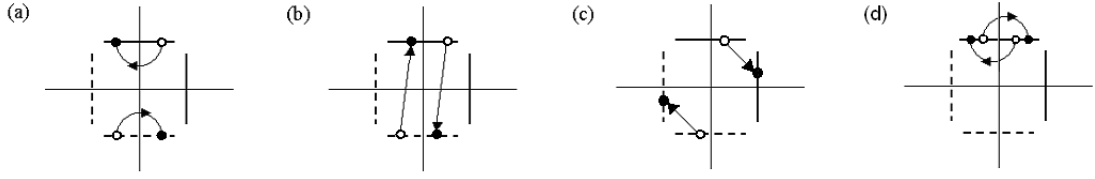}
\caption{\small The parametrization of the interactions in the
present model: (a) $U_{1R}$-processes, (b) $U_{2R}$-processes, (c)
$U_{3R}$-processes, and (d) $U_{4R}$-processes.} \label{f.2}
\end{figure}

Since our model is renormalizable, the counterterms originate
naturally from the initial form of the hamiltonian. Here we
neglect umklapp processes to start with. Then, the only divergent
terms that arise in perturbation theory come from the interaction
processes described by the $U_{1R}$ and $U_{2R}$ couplings as one
can most easily verify. In view of this, in a first order
approximation, we can disregard the perturbative terms coming from
the other two couplings, namely $U_{3R}$ and $U_{4R}$ and their
corresponding counterterms. As a result, the two sets of parallel
patches of FS decouple from each other, and there is no ambiguity
in locating a particle either at a solid line or at a dashed line
patch instead.

We begin by calculating first the electron self-energy in the vicinity
of $p_{\perp }=k_{FR}$. Using conventional Feynman rules, the diagrams
shown in fig.3 produce

\begin{eqnarray*}
\Sigma_{R}(p_{\perp},p_{\parallel})=\frac{\lambda\widetilde
U_{1R}}{4\pi^2}-(Z(\mathbf{p})-1)p_{0}+\frac{Z(\mathbf{p})}{2m_{B}}(k^{2}_{FR}-k^{2}_{FB})
\end{eqnarray*}

\begin{eqnarray*}
+(Z(\mathbf{p})\frac{m_{R}}{m_{B}}-1)v_{FR}(p_{\perp}-k_{FR})
-\frac{(3\Delta^{2}-p_{\parallel}^{2})}{256\pi^{4}\Delta^{2}
v_{FR}^{2}}(\widetilde U_{1R}^{2}+\widetilde
U_{2R}^{2})(p_{0}-v_{FR}(p_{\perp}-k_{FR}))
\end{eqnarray*}

\begin{eqnarray}
\times\left[\ln\left(\frac{\Omega-v_{FR}(p_{\perp}-k_{FR})-p_{0}-i\delta}{v_{FR}(p_{\perp}-k_{FR})-p_{0}-i\delta}\right)+
\ln\left(\frac{\Omega-v_{FR}(p_{\perp}-k_{FR})+p_{0}-i\delta}{v_{FR}(p_{\perp}-k_{FR})+p_{0}-i\delta}\right)\right]
\end{eqnarray}\\ \\
where $\widetilde{U}_{iR}(p_{\parallel })=\int dk_{\parallel
}U_{iR}(p_{\parallel },k_{\parallel })$. The parameter $\lambda $
is the ultraviolet momentum cutoff with the corresponding energy
cutoff given by $\Omega =2v_{FR}\lambda $, and $2\Delta $ is the
length of each path along $FS$. In general, the renormalized
coupling functions depend on three distinct momenta components
parallel to the Fermi surface. However, in calculating the Hartree
diagram, shown in fig.3(a), the vertex depend only on two
different momenta and we naturally arrive at the definition of
$\widetilde{U}_{iR}$. In contrast, in the sunset diagrams of
figs.3(b) and 3(c), the renormalized vertices depend explicitly on
three different momenta components along $FS$. In whatever way, in
two-loop order, the tadpole diagram is in fact the only
contribution from the self-energy which produce the
renormalization of the Fermi surface. As a result, for simplicity,
in a zeroth order approximation, we take
$\widetilde{U}_{iR}\left(p_{\Vert}\right)=(2\Delta)U_{iR}$,
neglect the momenta dependence of the vertices and rewrite the
renormalized coupling in terms of the corresponding
$\widetilde{U}_{iR}$'s. For comparison, we show in Appendix 1, how
the diagrams $\Sigma _{R}^{(3b)}$and $\Sigma _{R}^{(3c)}$ are
modified if the full momenta dependence of the renormalized
coupling functions are taken into account at all steps.

\begin{figure}
\includegraphics[height=1.1in]{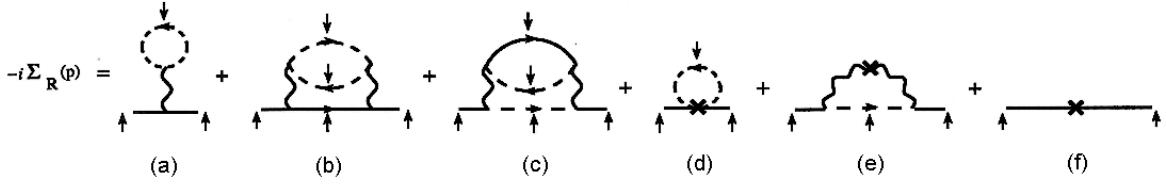}
\caption{\small The diagrams used for the calculation of the
renormalized self-energy. The ones with a cross represent the
counterterm contributions, which regularize the divergences.
Diagrams (d) and (e) cancel each other exactly. Diagram (f)
produces the contributions directly associated with Z.}
\label{f.3}
\end{figure}

In principle, since this a nonrelativistic system, there should be
two scaling parameters in the problem: one for the energy ($\omega
$) and another one for the momentum ($\Lambda $)~\cite{Kopietz}.
By making an appropriate choice of the renormalization
prescription, these two parameters do not mix with each other.
Here we choose to work with the energy scale $\omega $ only. In
doing this we implicitly assume that $\Lambda \rightarrow 0$ much
faster than $\omega \rightarrow 0$. Defining the renormalized
one-particle irreducible function $\Gamma _{R}^{(2)}(p_{\perp
},p_{\parallel },p_{0};\omega )$, which is nothing but the inverse
of the full single-particle Green's function, such that $Re\Gamma
_{R}^{(2)}(p_{\perp }=k_{FR},p_{\parallel },p_{0}=\omega ;\omega
\approx 0)=\omega $, we find that

\begin{equation}
Z(p_{\perp }=k_{FR},p_{\parallel };\omega \approx 0)=1-\frac{(3\Delta ^{2}-
p_{\parallel }^{2})}{128\pi ^{4}\Delta ^{2}v_{FR}^{2}}(\widetilde{U}_{1R}^{2}
+\widetilde{U}_{2R}^{2})\ln (\frac{\Omega }{\omega })+...,
\end{equation}
and

\begin{equation}
\mu _{B}=\frac{k_{FB}^{2}}{2m_{B}}=Z^{-1}(p_{\perp
}=k_{FR},p_{\parallel };\omega \approx 0)\left(\mu _{R}
+\frac{\lambda \widetilde{U}_{1R}}{4\pi ^{2}}\right)+...,
\end{equation}\\
where we assumed, for simplicity, that the bare and renormalized
masses differ only by a multiplicative factor, namely the charge
renormalization function, i.e., $m_{B}=Z(p_{\perp
}=k_{FR},p_{\parallel };\omega \approx 0)m_{R}$.

\section*{3 - Curvature and truncation in the renormalized $FS$}

Using perturbation theory, we calculate next the one-particle
irreducible functions $\Gamma
_{_{iR}}^{\left(4\right)}\left(p,k,q;\omega \right)$, which are
essentially the renormalized two-particle interaction. Since our
main intention in this work is to analyze the nature of the
resulting renormalized $FS$, we restrict ourselves to scattering
processes with two independent external momenta only. Employing
appropriate renormalization group prescriptions such as

\begin{equation}
\Gamma _{1R\uparrow \downarrow }^{\left(4\right)}\left(p_{\Vert },k_{\Vert },p_{0}+k_{0}=\omega ,k_{0}-p_{0}=\omega ;\omega \simeq 0\right)=-iU_{1R}\left(p_{\Vert },k_{\Vert };\omega \right),\label{eq:}\end{equation}

\begin{equation}
\Gamma _{2R\uparrow \downarrow }^{\left(4\right)}\left(p_{\Vert },k_{\Vert },p_{0}+k_{0}=\omega ;\omega \simeq 0\right)=-iU_{2R}\left(p_{\Vert },k_{\Vert };\omega \right),\label{eq:}\end{equation}
and

\begin{equation}
\Gamma _{2R\uparrow \uparrow \left(\downarrow \downarrow
\right)}^{\left(4\right)}\left(p_{\Vert },k_{\Vert
},p_{0}+k_{0}=\omega ;\omega \simeq
0\right)=0,\label{eq:}\end{equation}\\
for $p_{\bot }=-k_{\bot }=k_{FR}$, together with the perturbative
expansions for the one-particle irreducible functions, we can
relate the corresponding renormalized and bare coupling functions
to each other. Following the same approximating scheme as before,
with respect the dependence on the momentum component parallel to
$FS$ in the vertices of the various Feynman diagrams, and taking
into account the RG conditions $\omega \partial
\widetilde{U}_{iB}/\partial \omega =0$, we find the resulting RG
equations

\begin{equation}
\beta _{\widetilde{U}_{1R}}=\frac{1}{v_{FR}}\omega \frac{\partial \widetilde{U}_{1R}}{\partial \omega }=
\frac{\left(3\Delta ^{2}-p_{\Vert }^{2}\right)}{16\pi ^{2}v_{FR}^{2}\Delta ^{2}}\widetilde{U}_{2R}^{2}
+\frac{\left(17\Delta ^{2}-3p_{\Vert }^{2}\right)}{384\pi ^{4}v_{FR}^{3}\Delta ^{2}}
\widetilde{U}_{1R}\left(\widetilde{U}_{1R}^{2}+\widetilde{U}_{2R}^{2}\right)+...,\label{eq:}
\end{equation}

\begin{equation}
\beta _{\widetilde{U}_{2R}}=\frac{1}{v_{FR}}\omega \frac{\partial \widetilde{U}_{2R}}{\partial \omega }=
\frac{\left(3\Delta ^{2}-p_{\Vert }^{2}\right)}{8\pi ^{2}v_{FR}^{2}\Delta ^{2}}\widetilde{U}_{2R}
\widetilde{U}_{1R}+\frac{\left(17\Delta ^{2}-3p_{\Vert }^{2}\right)}{384\pi ^{4}v_{FR}^{3}\Delta ^{2}}
\widetilde{U}_{2R}\left(\widetilde{U}_{1R}^{2}+\widetilde{U}_{2R}^{2}\right)+....\label{eq:}
\end{equation}\\
For comparison, we show in Appendix 2 how the RG equations for the
corresponding renormalized coupling functions look like if we take
into account the full momenta dependence at all stages. Finally,
using eq.(5), and following a similar procedure with the
renormalized Fermi energy $\mu _{R}\left(p_{\Vert },\omega
\right)$, we get the last RG equation of our interest

\begin{equation}
\beta _{\mu _{R}}=\frac{1}{\Omega }\omega \frac{\partial \mu _{R}}{\partial \omega }=\frac{\left(3\Delta ^{2}
-p_{\Vert }^{2}\right)}{128\pi ^{4}v_{FR}^{2}\Delta ^{2}\Omega }\left(\widetilde{U}_{1R}^{2}
+\widetilde{U}_{2R}^{2}\right)\left(\mu _{R}+\frac{\Omega \widetilde{U}_{1R}}{8\pi ^{2}v_{FR}}\right)
-\frac{1}{8\pi ^{2}}\omega \frac{\partial }{\partial \omega }\left(\frac{\widetilde{U}_{1R}}{v_{FR}}\right)
+....\label{eq:}
\end{equation}

\begin{figure}
\centering \includegraphics[height=2.7in]{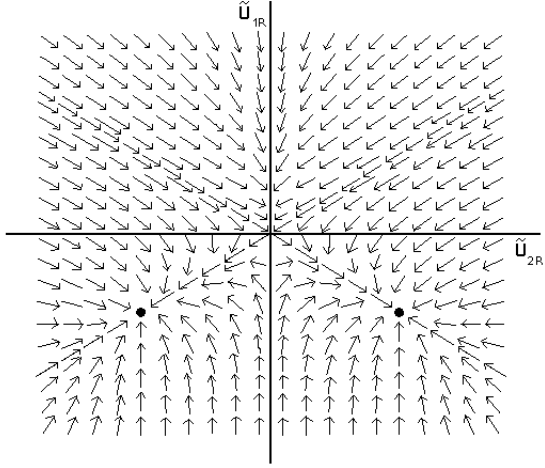} \caption{\small
The flow diagram in the ($\widetilde U_{1R},\widetilde U_{2R}$)
plane.} \label{f.4}
\end{figure}

To determine the fixed points of the model, we calculate next the
zeros of these RG equations for both renormalized couplings, and
the renormalized Fermi energy. It then turns out that, aside from
the usual infrared unstable Fermi liquid fixed point, and yet
another nontrivial unstable fixed point, we get two infrared
stable nontrivial fixed points which are, as we shall see,
associated with non-Fermi liquid phases. For conciseness, we will
only present the final expression for these IR stable fixed points

\begin{equation}
\widetilde{U}_{1R}^{*}=-16\pi ^{2}v_{FR}^{*}\left(\frac{3\Delta ^{2}-p_{\parallel }^{2}}{17\Delta ^{2}-3p_{\parallel }^{2}}\right)\end{equation}

\begin{equation}
\widetilde{U}_{2R}^{*}=\pm \sqrt{2}\widetilde{U}_{1R}^{*}\end{equation}

\begin{equation}
k_{FR}^{*}=8\lambda \left(\frac{3\Delta ^{2}-p_{\parallel }^{2}}{17\Delta ^{2}-3p_{\parallel }^{2}}\right)
\end{equation}\\ \\
We observe that they depend upon $p_{\parallel }$ in a essential
way. In view of that, the Fermi surface of the system also
acquires a $p_{\parallel }$-structure and deviates slightly from
its initial flat form. This FS deformation comes out naturally
from the renormalization process. This feature of the fixed points
appears only in calculations up to two-loop order or beyond. In
fig.4, we show schematically the scaling trajectories in the
($\widetilde{U}_{1R},\widetilde{U}_{2R}$) plane.

The nature of the electron liquid associated with the nontrivial
fixed points can be inferred by the flow of the charge
renormalization function $Z(p_{\parallel },\omega )$. Since
$\gamma =(\omega \slash Z)(\partial Z\slash \partial \omega )$, in
the vicinity of the fixed point, we have that $Z(p_{\parallel
},\omega )$ scales as $(\omega \slash \Omega )^{\gamma
^{*}(p_{\parallel })}$ with the anomalous dimension being simply
$\gamma $ evaluated at those fixed point values. Note that it is
always positive definite along FS. Calculating $\gamma
^{*}(p_{\parallel })$ explicitly, we find that, indeed, the charge
renormalization function vanishes most rapidly at the center of
the FS patch (fig.5). This anisotropic suppression of
$Z(p_{\parallel },\omega )$ was also emphasized by Kishine and
Yonemitsu within a Wilsonian RG approach~\cite{Kishine}.

\begin{figure}
\centering \includegraphics[height=2.7in]{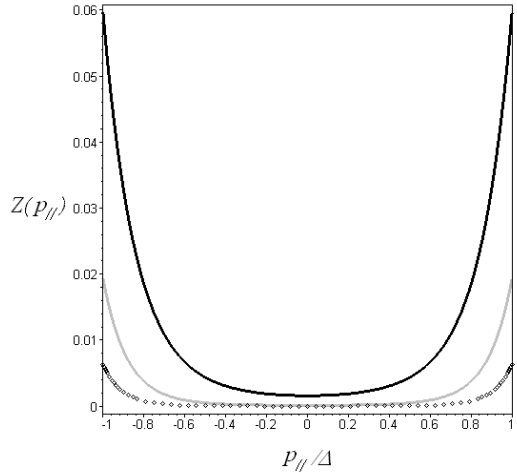}
\caption{\small The anisotropic suppression of $Z(p_{\parallel})$
in the present model as we approach the FS for three values of
$(\omega\slash\Omega)$. The black line is for
$(\omega\slash\Omega)=10^{-5}$, the light grey line is for
$(\omega\slash\Omega)=10^{-7}$, and the dotted line is for
$(\omega\slash\Omega)=10^{-9}$.} \label{f.5}
\end{figure}

Another particularly interesting feature is that, although the FS
is IR stable within given boundaries, there is no guarantee that the
physical FS is well-defined throughout the original patch. The nature
of the resulting fermionic system can be determined by the behavior
of the corresponding momentum distribution function $n(\mathbf{p})$.
If $n(\mathbf{p})$ has an infinite slope at the renormalized FS,
the state is metallic and resembles a Luttinger liquid. If it turns
out otherwise that $n(\mathbf{p})$ is perfectly smooth at $k_{FR}$,
the renormalized FS is truncated, and there appears a charge pseudogap
at those points. This gapped state depicts an insulating spin liquid
instead. To make the argument more quantitative, we use the momentum
distribution function calculated for a similar FS in ref.~\cite{Ferraz2}.
In that work, it was shown explicitly that, for $1/2<\gamma ^{*}(p_{\parallel })<1$,
the FS must be truncated at the corresponding $p_{\parallel }$ values.
Here this condition is fulfilled for $|(p_{\parallel }\slash \Delta )|\leq 0.41$.
After the elimination of the corresponding Fermi surface segments,
the remains of the interacting FS are shown in fig.6. This truncation
scenario will be discussed in great length elsewhere~\cite{Hermann}.

\begin{figure}
\centering \includegraphics[height=2.6in]{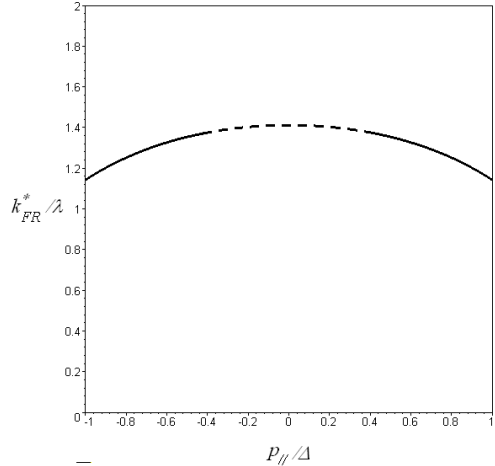}
\caption{\small The renormalized Fermi surface of the present
model. The dashed line stands for truncation.} \label{f.6}
\end{figure}

\section*{4 - Conclusion}

In summary, we showed explicitly by a two-loop RG calculation
that, even if the initial FS is entirely flat in two spatial
dimensions, the true interacting Fermi surface becomes slightly
curved as a result of interactions. Using field theoretical
methods, we explored the existence of infrared stable nontrivial
fixed points, which are associated with the non-Fermi liquid
behavior in the low-energy sector. We showed that the fixed points
vary continuously with the momentum $p_{\parallel }$ along the
Fermi surface. In addition, we argued that the criterion $Z(\omega
)\rightarrow 0$ as $\omega \rightarrow 0$ does not suffice in
determining the nature of the resulting non-Fermi liquid state. We
called attention to a possible route of physical characterization
of those states in terms of the momentum distribution function.
For the $p_{\parallel }$ values, in which the anomalous dimension
$\gamma (p_{\parallel })$ is such that $1/2\leq \gamma
^{*}(p_{\parallel })\leq 1$, the renormalized FS suffers a
truncation. This takes place, in our case, at the central region
of the original FS patches.

Not long ago, a new ARPES data for the high-temperature
superconductors $Bi_{2}Sr_{2-x}La_{x}CuO_{6-\delta}$ was reported
showing some indication of spin-charge separation and
``Luttinger'' liquid like behavior in the normal state of that
compound~\cite{Janowitz}. That data was, most recently, fitted
consistently by a ``Luttinger'' liquid-like
phenomenology~\cite{Byczuk}. Those workers found that, contrary to
what happens in one dimension, the anomalous exponents vary
strongly with momentum along the Fermi surface. This is in
agreement with the findings of our work. We also note that, in the
cuprates, there exists a pseudogap phase with a truncated FS
around special k-values in momentum space. In view of that, it is
therefore natural to ascribe to those gapped systems an insulating
spin liquid nature instead~\cite{Rice}. Again, it is suggestive to
associate that result to our work. It is therefore important to
explore this scenario further. In a even more general experimental
framework, one could also try to apply pressure in a interacting
metallic state to move the Fermi surface towards its critical
condition inducing, in this way, a new kind of quantum phase
transition~\cite{Aepli}. This would, certainly, open more
possibilities to test the limits of our results.\\

\emph{Acknowledgements} - This work was partially supported by the
Financiadora de Estudos e Projetos (FINEP) and by the Conselho
Nacional de Desenvolvimento Científico e Tecnológico (CNPq) -
Brazil.

\section*{Appendix 1}

Using conventional Feynman rules, the contributions of the
diagrams from figs.3(b) and 3(c) for the self-energy and the
non-interacting single-particle Green's functions associated with
our model hamiltonian are

\begin{eqnarray*}
\Sigma _{R}^{\left(3b\right)}\left(p_{\Vert },p_{\bot
}=k_{FR}\right)=-\frac{1}{64\pi ^{4}}\int _{D_{3}} dk_{\Vert
}dq_{\Vert }\left(\frac{1}{v_{FR}^{2}}\right)U_{1R}\left(k_{\Vert
},-k_{\Vert }+p_{\Vert }+q_{\Vert },q_{\Vert }\right)
\end{eqnarray*}

\begin{eqnarray*}
\times U_{1R}\left(p_{\Vert },q_{\Vert },-k_{\Vert }+p_{\Vert
}+q_{\Vert }\right) \left(p_{0}-v_{FR}(p_{\perp }-k_{FR})\right)
\end{eqnarray*}

\begin{equation}
\times\left[\ln \left(\frac{\Omega -v_{FR}(p_{\perp
}-k_{FR})-p_{0}-i\delta }{v_{FR}(p_{\perp }-k_{FR}) -p_{0}-i\delta
}\right)+\ln \left(\frac{\Omega -v_{FR}(p_{\perp
}-k_{FR})+p_{0}-i\delta }{v_{FR}(p_{\perp } -k_{FR})+p_{0}-i\delta
}\right)\right]
\end{equation}\\
and

\begin{eqnarray*}
\Sigma _{R}^{\left(3c\right)}\left(p_{\Vert },p_{\bot
}=k_{FR}\right)=-\frac{1}{64\pi ^{4}}\int _{D_{3}}dk_{\Vert
}dq_{\Vert }\left(\frac{1}{v_{FR}^{2}}\right)U_{2R}\left(-k_{\Vert
}+p_{\Vert }+q_{\Vert },k_{\Vert },q_{\Vert }\right)
\end{eqnarray*}

\begin{eqnarray*}
\times U_{2R}\left(p_{\Vert },q_{\Vert },k_{\Vert }\right)
\left(p_{0}-v_{FR}(p_{\perp }-k_{FR})\right)
\end{eqnarray*}

\begin{eqnarray}
\times\left[\ln \left(\frac{\Omega -v_{FR}(p_{\perp
}-k_{FR})-p_{0}-i\delta }{v_{FR}(p_{\perp }-k_{FR}) -p_{0}-i\delta
}\right)+\ln \left(\frac{\Omega -v_{FR}(p_{\perp
}-k_{FR})+p_{0}-i\delta }{v_{FR}(p_{\perp } -k_{FR})+p_{0}-i\delta
}\right)\right]
\end{eqnarray}\\
where the domain of integration $D_{3}$ is $-\Delta \leq k_{\Vert
}\leq \Delta $, $-\Delta \leq q_{\Vert }\leq \Delta $, and
$-\Delta \leq -k_{\Vert }+p_{\Vert }+q_{\Vert }\leq \Delta $. If,
for simplicity, the coupling functions and the Fermi velocity are
considered to be independent of the momenta components parallel to
$FS$, our results follow immediately.

\section*{Appendix 2}

Using perturbation theory together with our renormalization group
prescriptions, and taking full account of the coupling functions
dependence on the momenta components parallel to $FS$ the RG
equations for $U_{1R}$ and $U_{2R}$, for general scattering
processes, become

\begin{eqnarray*}
\omega \frac{\partial U_{1R}\left(p_{1\Vert },p_{2\Vert
},p_{3\Vert }\right)}{\partial \omega } =\frac{1}{4\pi
^{2}}\bigg\{ \int _{D_{1}}dk_{\Vert
}\left(\frac{1}{v_{FR}}\right)\bigg[U_{1R} \left(p_{1\Vert
},p_{2\Vert },k_{\Vert }\right)U_{1R}\left(p_{1\Vert }+p_{2\Vert
}-k_{\Vert },k_{\Vert } ,p_{3\Vert }\right)
\end{eqnarray*}

\begin{eqnarray*}
+U_{2R}\left(p_{1\Vert }, p_{2\Vert },k_{\Vert
}\right)U_{2R}\left(p_{1\Vert }+p_{2\Vert }-k_{\Vert },k_{\Vert
},p_{3\Vert }\right)\bigg]-\int _{D_{2}}dk_{\Vert
}\left(\frac{1}{v_{FR}}\right)\bigg[U_{1R}\left(p_{1\Vert },
p_{3\Vert}-p_{1\Vert}+k_{\Vert },p_{3\Vert }\right)
\end{eqnarray*}

\begin{eqnarray*}
\times U_{1R}\left(k_{\Vert },p_{2\Vert },p_{3\Vert }-p_{1\Vert
}+k_{\Vert }\right)\bigg]\bigg\} +\frac{1}{64\pi
^{4}}U_{1R}\left(p_{1\Vert },p_{2\Vert },p_{3\Vert }\right)
\end{eqnarray*}

\begin{eqnarray*}
\times\sum
_{i=1}^{4}\delta_{p_{1\Vert}+p_{2\Vert},p_{3\Vert}+p_{4\Vert}}\int
_{D_{3}}dk_{\Vert }dq_{\Vert
}\left(\frac{1}{v_{FR}^{2}}\right)\bigg[U_{1R}\left(p_{i\Vert
},q_{\Vert },-k_{\Vert}+p_{i\Vert }+q_{\Vert }\right)
\end{eqnarray*}

\begin{eqnarray}
\times U_{1R}\left(k_{\Vert },-k_{\Vert }+p_{i\Vert }+q_{\Vert
},q_{\Vert }\right)+U_{2R}\left(p_{i\Vert },q_{\Vert },k_{\Vert
}\right)U_{2R}\left(-k_{\Vert }+p_{i\Vert } +q_{\Vert },k_{\Vert
},q_{\Vert }\right)\bigg]\label{eq:}
\end{eqnarray}
and

\begin{eqnarray*}
\omega \frac{\partial U_{2R}\left(p_{1\Vert },p_{2\Vert
},p_{3\Vert }\right)}{\partial \omega } =\frac{1}{4\pi
^{2}}\bigg\{ \int _{D_{1}}dk_{\Vert
}\left(\frac{1}{v_{FR}}\right)\bigg[U_{1R}\left(p_{1\Vert },
p_{2\Vert },k_{\Vert
}\right)U_{2R}\left(p_{1\Vert}+p_{2\Vert}-k_{\Vert},k_{\Vert},p_{3\Vert}\right)
\end{eqnarray*}

\begin{eqnarray*}
+U_{1R}\left(p_{1\Vert }+p_{2\Vert }-k_{\Vert },k_{\Vert
},p_{3\Vert }\right)U_{2R}
\left(p_{1\Vert},p_{2\Vert},k_{\Vert}\right)\bigg]\bigg\}
+\frac{1}{64\pi ^{4}}U_{2R}\left(p_{1\Vert },p_{2\Vert },p_{3\Vert
}\right)
\end{eqnarray*}

\begin{eqnarray*}
\times\sum
_{i=1}^{4}\delta_{p_{1\Vert}+p_{2\Vert},p_{3\Vert}+p_{4\Vert}}\int
_{D_{3}}dk_{\Vert }dq_{\Vert
}\left(\frac{1}{v_{FR}^{2}}\right)\bigg[U_{1R}\left(p_{i\Vert
},q_{\Vert },-k_{\Vert}+p_{i\Vert }+q_{\Vert }\right)
\end{eqnarray*}

\begin{eqnarray}
\times U_{1R}\left(k_{\Vert },-k_{\Vert }+p_{i\Vert }+q_{\Vert
},q_{\Vert }\right)+U_{2R}\left(p_{i\Vert },q_{\Vert },k_{\Vert
}\right)U_{2R}\left(-k_{\Vert }+p_{i\Vert } +q_{\Vert },k_{\Vert
},q_{\Vert }\right)\bigg]\label{eq:}
\end{eqnarray}
where the domains $D_{1}$, $D_{2}$ and $D_{3}$ are given by

\begin{displaymath}
D_{1} = \left\{ \begin{array}{ll} -\Delta
\leq k_{\Vert }\leq \Delta\\
-\Delta \leq p_{1\Vert }+p_{2\Vert
}-k_{\Vert }\leq \Delta\\
\end{array} \right.
\end{displaymath}

\begin{displaymath}
D_{2} = \left\{ \begin{array}{ll} -\Delta
\leq k_{\Vert }\leq \Delta\\
-\Delta \leq p_{3\Vert }-p_{1\Vert
}+k_{\Vert }\leq \Delta\\
\end{array} \right.
\end{displaymath}

\begin{displaymath}
D_{3} = \left\{ \begin{array}{ll} -\Delta
\leq k_{\Vert }\leq \Delta\\
-\Delta \leq q_{\Vert }\leq \Delta\\
-\Delta \leq p_{i\Vert }+q_{\Vert}-k_{\Vert }\leq \Delta\\
\end{array} \right.
\end{displaymath}\\
Once again if, for simplicity, we neglect the the dependence of
the coupling functions with respect the components of the momenta
along $FS$, consider scattering processes associated with only two
independent external momenta and integrate over one of them, our
simplified RG equations for the $\widetilde{U}_{iR}$'s are readily
reproduced.


\begin{thebibliography}{10}
\bibitem{Anderson}P.W. Anderson, Science \textbf{235}, 1196
(1987);
\bibitem{Campuzano}See, e.g., J. C. Campuzano, M. R. Norman and M. Randeria, \emph{Physics
of Conventional and Unconventional Superconductors}, edited by K.
H. Benneman and J. B. Ketterson, Springer-Verlag (2003);
\bibitem{Dessau}D. S. Dessau et al, Phys. Rev. Lett. \textbf{71}, 2781 (1993);
\bibitem{Yoshida}T. Yoshida et al, Phys. Rev. Lett. \textbf{91}, 27001 (2003);
\bibitem{Ruvalds}A. Virosztek and J. Ruvalds, Phys. Rev B \textbf{42}, 4064 (1990);
\bibitem{Dzyaloshinskii}A. T. Zheleznyak, V. M. Yakovenko and I. E. Dzyaloshinskii, Phys.
Rev B \textbf{55}, 3200 (1997);
\bibitem{Doucot}F. V. Abreu and B. Douçot, Europhys. Lett. \textbf{38}, 533 (1997);
\bibitem{Luther}A. Luther, Phys. Rev. B \textbf{50}, 11446 (1994);
\bibitem{Sudbo}J. O. Fjaerestad, A. Sudbo and A. Luther, Phys. Rev. B \textbf{60}, 13361 (1999);
\bibitem{Nozieres}P. Nozieres, \emph{Interacting Fermi Systems}, Benjamim Press (1964);
\bibitem{Solyom}J. Solyom, Adv. Phys. \textbf{28}, 202 (1979);
\bibitem{Ferraz1}A. Ferraz, Europhys. Lett. \textbf{61}, 228 (2003);
\bibitem{Ferraz2}A. Ferraz, Phys. Rev B \textbf{68}, 75115 (2003);
\bibitem{Peskin}M. E. Peskin and D. M. Schroeder, \emph{An Introduction to Quantum Field
Theory}, Perseus Books (1995);
\bibitem{Kopietz}P. Kopietz and T. Busche, Phys. Rev. B \textbf{64}, 155101 (2001);
\bibitem{Kishine}J. Kishine and K. Yonemitsu, Phys. Rev. B \textbf{59}, 14823 (1999);
\bibitem{Hermann}H. Freire and A. Ferraz, \emph{in preparation};
\bibitem{Janowitz}C. Janowitz, R. Müller, L. Dudy, A. Krapf, R. Manzke, C. R. Ast and H. Höchst,
Europhys. Lett. \textbf{60}, 615 (2002);
\bibitem{Byczuk}K. Byczuk, C. Janowitz, R. Müller, R. Manzke, J.
Spalek and W. Wójcik, cond-mat/0405552, to appear in Europhys
Lett. (2004);
\bibitem{Aepli}G. Aeppli, \emph{private communication};
\bibitem{Rice}N. Furukawa, T. M. Rice and M. Salmhofer, Phys. Rev. Lett. \textbf{81}, 3195
(1998).
\end{thebibliography}
\end{document}